# SIMULATIONS OF CLUSTERS OF GALAXIES

*August E. Evrard*

Department of Physics, University of Michigan, Ann Arbor, MI 48109-1120 USA

**Abstract**

The degree of complexity and, to a somewhat lesser degree, realism in simulations has advanced rapidly in the past few years. The simplest approach — modeling a cluster as collisionless dark matter and collisonal, non–radiative gas is now fairly well established. One of the most fruitful results of this approach is the *morphology–cosmology connection* for X–ray clusters. Simulations have provided the means to make concrete predictions for the X–ray morphologies of clusters in cosmologies with different $\Omega_o$, with the result that low $\Omega_o$ cosmologies fair rather poorly when compared to observations. Another result concerns the accuracy of X–ray binding mass estimates. The standard, hydrostatic, isothermal model estimator is found to be accurate to typically better than 50% at radii where the density contrast is between $10^2$ and $10^3$.

More complicated approaches, which attempt to explicitly follow galaxy formation within the proto–cluster environment are slowly being realized. The key issue of *dynamical biasing* of the galaxy population within a cluster is being probed, but no conclusive understanding has been realized. The dynamics of multi–phase gas, including conversion of cold, dense gas into stars and the feedback therefrom, is the largest obstacle hindering progress. An example demonstrating the state–of–the–art in this area is presented.

## 1. Introduction

A typical rich cluster is a multi–component system containing many tens to hundreds of bright galaxies, a hot, metal–enriched intracluster medium (ICM) observed in X–rays, and dark matter whose presence has been inferred by application of the virial theorem for over 60 years [27]. The present relative distributions of these components reflect their full dynamical and thermal histories, which need not be the same. The role of simulations is to provide a tool to investigate the dynamical evolution of cluster components in a cosmological setting. A unique value of this tool is the ability to 'synthetically image' the results in a well–prescribed manner, providing direct, 'apples–to–apples' comparisons between theory and observation.

Nearly all viable large-scale structure models are 'bottom–up', in the sense that collapse of galactic–sized perturbations (= galaxy formation?) precedes the collapse of cluster–sized perturbations. In this way, aspects of cluster formation are intimately tied to galaxy formation which, of course, is linked to cooling and fragmentation of gas clouds and star formation [26]. Many unsettled issues regarding clusters (*e.g.*, the origin and distribution of metals in the ICM, whether or not galaxies fairly trace the cluster dark matter) persist because of the uncertainties in modeling, from both a physical and numerical perspective, galactic–scale star formation.

One can exploit the 'bottom–up' picture to simplify the problem considerably by assuming that the star formation within galaxies is largely finished before the collapse of the bulk of the cluster. One then thinks of the intracluster medium as the leftovers of galaxy formation which simply fall into and shock heat within the dominant, cluster potential well. This 'primordial infall' hypothesis, introduced by Gunn & Gott [12], has empirical reinforcement in the fact that the baryon content of the largest clusters is dominated by the intracluster gas rather than the galaxies [3]. Below, I present results using this approach applied to the problems of using the X–ray morphology of clusters to constrain the density parameter $\Omega_o$ (§3) and to the problem of the reliability of X–ray based binding mass estimates (§4).

Ideally, one would like to follow the formation of the galaxies directly within the proto–cluster environment. Physically, this requires adding, at a minimum, the dissipative mechanism of radiative cooling within the code. Physics associated with star formation can also be included, although with a large degree of parameterization uncertainty. Numerically, a wide dynamic range and enhanced spatial resolution are required since individual galaxies are factors $\gtrsim 10^3$ less massive than clusters and the optically bright regions of galaxies represent density enhancements a factor $\gtrsim 10^6$ over the cosmological background.

Some of the work presented in this article, in particular, the projects in §3 and §5, has been written up in another recent review [5]. There are recent developments in both projects which are included here. The work in §4 is new. Where needed, $h = 0.5$ is assumed throughout, with $h \equiv H_o/100 \text{ km s}^{-1} \text{ Mpc}^{-1}$.

## 2. Simulation Flavors

Simulations now come in a variety of flavors, as summarized in Fig. 1. Advances in the late 70's to mid 80's were made using N–body simulations which model the gravitationally dominant dark matter component. This approach is limited by its inability to follow directly observable components. One must make assumptions, such as galaxies fairly tracing the mass, in order to make contact with observations. The late 80's saw the advent of coupling gas dynamics with N–body codes, providing the capability to directly follow the baryonic component coupled to, but independent of, the dark matter. Multi–fluid approaches, incorporating several dynamically distinct components represent the current state–of–the–art.

In modeling the gas dynamics, one must choose whether to employ an Eulerian or Lagrangian approach. The simulations discussed in this article all use the Lagrangian method of smoothed particle hydrodynamics (SPH) [18, 13] coupled to the P3M N–body code[6]. Details of the P3MSPH can be found in ref [4]. Briefly, SPH is a Lagrangian scheme which uses a smoothing kernel $W(r, h)$ to determine characteristics of the fluid at a given point based on properties associated with the local particle distribution. For example, the density at the position of particle $i$ is given by

$$\rho_i = \sum_j m_j \, W(r_{ij}, h_i) \qquad (1)$$

where $r_{ij}$ is the separation of the pair of particles $i$ and $j$ and $h_i$ is the local smoothing scale. The kernel $W$ has compact support on a scale of a few $h$; hence, $h$ is a measure of the local resolution of the solution. Usually $h$ is adaptively varied both spatially and temporally such that a constant number of neighbors in the range $\mathcal{O}(10^2)$ is involved in the above sum. This extends the dynamic range available in the experiment.

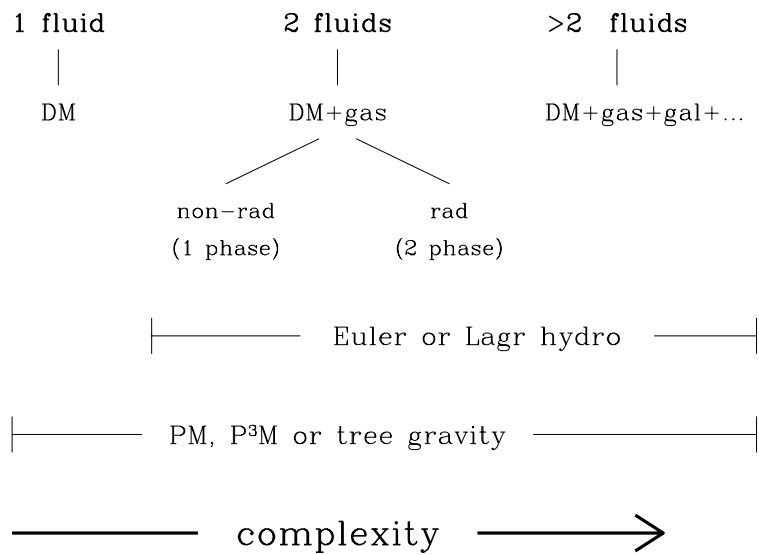

Figure 1. Schematic showing the variety of simulation approaches available for modeling structure formation.

The Lagrangian nature and wide dynamic range of SPH are well suited to the problem of large–scale structure formation. Schemes using Eulerian finite difference methods with fixed spatial resolution are limited in their ability to resolve objects of very high density contrast (such as galaxies) within a large–scale cosmological environment. Conversely, Lagrangian codes which employ particles of fixed mass have difficulty resolving low density regions where, by definition, there is not much mass.

Consider a region of space modeled by either $N_{part}$ Lagrangian particles or $N_{cell}$ Eulerian cells. Roughly speaking, the 'break–even' density contrast $\delta_{eq}$, where Eulerian and Lagrangian approaches have comparable resolution, is that at which one particle in the Lagrangian calculation is contained in one cell of the Eulerian code. At densities above $\delta_{eq}$, the Lagrangian method resolves one Eulerian cell with more than one particle (implying 'higher resolution') while for densities less than $\delta_{eq}$, a single Lagrangian particle covers many Eulerian cells ('lower resolution'). It follows then that the 'break–even' density contrast is $\delta_{eq} = N_{cell}/N_{part}$.

In three dimensions, $\delta_{eq} = 256^3/64^3 = 64$ is a presently realistic value. A comparison of several cosmological gas dynamic schemes applied to structure formation in a cold dark matter universe supports this simple argument [14]. Examined at low spatial resolution, the codes produce very similar results for the thermal and spatial structure of the gas. When examined on smaller spatial scales, the Eulerian codes display superior resolution in low density contrast regions, whereas higher resolution is achieved by the SPH codes in high density contrast regions. An example of the latter is shown in Fig. 2. Since the mass within an Abell radius of rich clusters represents a significant ($\delta > 100$) local density enhancement, a Lagrangian approach is (arguably, of course) currently the most efficient and effective means to model them numerically. When estimating X–ray luminosities of clusters, resolution of high density regions is essential since the emission measure of bremsstrahlung scales as $\rho^2$.

## 3. A Morphology–Cosmology Connection

There are several ways clusters can be used as cosmological diagnostics (see West, White, Kaiser, Rhee and others in this proceedings). Their abundance as a function of, for example, velocity dispersion $\sigma$, is extremely sensitive to the normalization of the fluctuation spectrum (see den Hartog in this proceedings). The dependence of the shape of cluster density profiles to cosmology has also been recently re–examined [2] and is a potentially useful cosmological diagnostic.

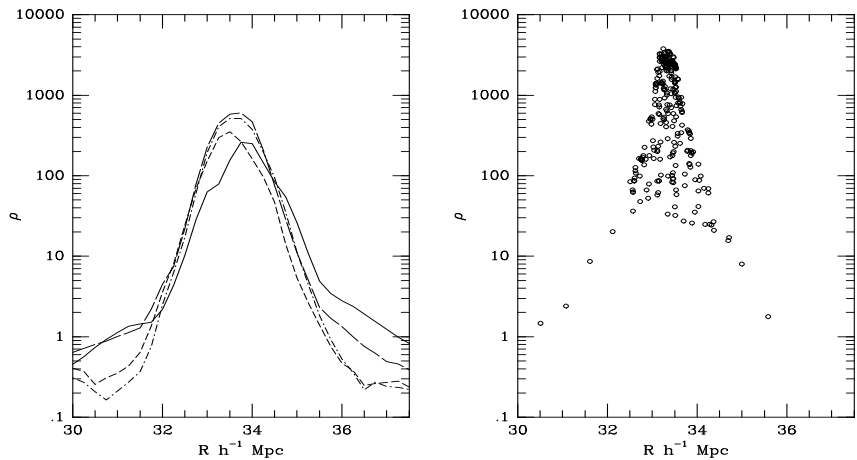

Figure 2. Density as a function of position along a rectangular region containing one of the larger clusters in the test simulation of Kang *et al.* (1994). The P3MSPH solution, in the right panel, under–resolves the low density 'wings' surrounding the cluster, where there is little mass and, hence, few particles. Conversely, the Eulerian code, shown in the left panel, under–resolves the central, condensed part of the cluster, where the bulk of the mass resides (as shown by the particles in the right panel). The density underestimate in the Eulerian code is due to a lack of spatial resolution; much of the cluster mass lies within one grid cell. The adaptive nature of the SPH smoothing kernel provides higher resolution in dense regions.

A new approach to constraining $\Omega_o$ is based on the structure of the hot, intracluster gas. The motivating idea, pointed out by Richstone, Loeb & Turner [22], is that, because the linear growth of perturbations diminishes as $\Omega$ decreases, structure formation in a low $\Omega_o$ universe should occur earlier than if $\Omega = 1$. Their analysis based on a spherical model for cluster collapse yields an age difference between clusters in models with $\Omega_o = 0.2$ and $\Omega = 1$ of $\sim 0.3\, H_o^{-1} \sim 4 - 6$ billion years. The sound crossing time for 10 keV gas in the central 1 Mpc of a cluster is only 0.6 billion years, so this age difference corresponds to many sound crossing times within the region surveyed by X–ray imaging instruments. This leads to the expectation that clusters in low density models should have more relaxed X–ray isophotes than their critical counterparts.

This effect has now been verified and quantified with a set of 24 P3MSPH simulations [8]. Eight sets of initial conditions, two each in comoving periodic boxes of side 30, 40, 50 and 60 Mpc, were generated in a constrained manner [1] from an initial CDM fluctuation spectrum. Each initial density field was evolved in three different cosmological backgrounds: (i) an unbiased, open universe with $\Omega_o = 0.2$; (ii) an unbiased, vacuum energy dominated universe with $\Omega_o = 0.2$ and $\lambda_o = 0.8$ and (iii) a biased, critical density ($\Omega = 1$) universe with *rms* present, linear mass fluctuations in a sphere of $8\, h^{-1}$ Mpc equal to $\sigma_8 = 0.59$. A baryon content of $\Omega_b = 0.1$ was assumed for the models, with all the baryons in the form of gas. The rest of the mass was assumed to be collisionless dark matter.

Examples of present day X–ray images of the simulated clusters are shown in Fig. 3, along with corresponding *Einstein* IPC images of real Abell clusters. The simulated images show the IPC band–limited flux in a 68′ square field with an angular resolution of about 2′. The simulations were placed at the same redshift, scaled to have the same X–ray temperature and 'observed' for the same amount of time as the real clusters. Realistic detector noise was added. The low density models, shown in the last two rows, are much more centrally concentrated and display much less asymmetry than the critical universe clusters shown in the second row. These differences arise because the low $\Omega_o$ clusters suffer fewer merging events at late times, a result expected from a variety of analytic

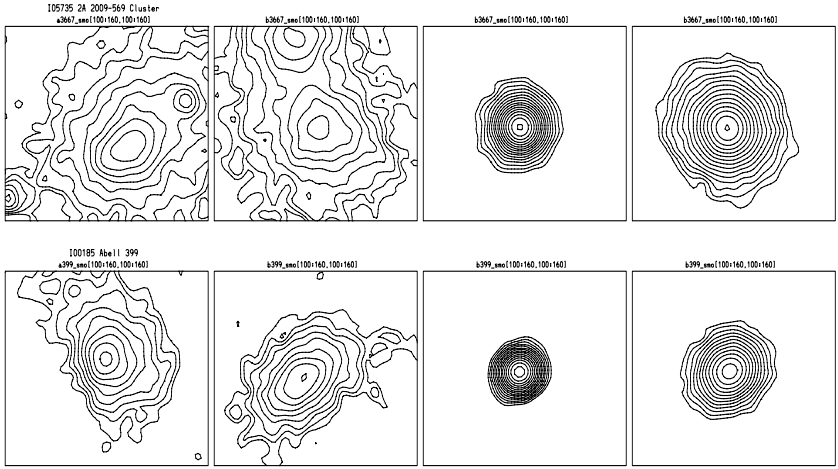

Figure 3. Examples of X–ray images for real and simulated clusters. The left column shows *Einstein* IPC archive images of Abell 3667 and 399 (upper and lower, respectively). The next three columns show randomly oriented views of randomly selected simulated clusters placed at the same redshift, scaled to the same X–ray temperature, and 'viewed' using the appropriate IPC instrument response for the same amount of time as the real clusters. Instrument noise (but no background sources) was introduced to the simulated images. Starting from the second to left, the columns are for the following models: $\Omega = 1$, $\Omega_o = 0.2$ with $\lambda = 0$ and $\Omega_o = 0.2$ with $\lambda = 0.8$. A CDM spectrum with $\Gamma = 0.5$ was used to generate the initial conditions.

arguments [22, 17, 15].

We have quantified the differences using statistics measuring the surface brightness fall-off (the familiar $\beta_{fit}$ parameter), mean isophotal center shift [8, 21] and the mean eccentricity. The same measures have been made for both the simulated and observed clusters. Histograms of any of these show a distinct difference between the low and high density models, with the observations strongly favoring $\Omega = 1$ over either of the $\Omega_o = 0.2$ universes [20]. This result is supported by recent analysis of the abundance of rich clusters. In order to reproduce observations, an $\Omega_o = 0.2$ CDM dominated universe requires a very high fluctuation amplitude $\sigma_8 = 1.25 - 1.58$, which requires galaxies to be *less* clustered than the mass distribution [24].

Although there is more physics beyond this simple approach, it is difficult to finger a mechanism which would strongly distort in an *anisotropic fashion* the present cluster X–ray morphologies in the low $\Omega_o$ models. Feedback due to winds from early–type galaxies would have to occur very recently, and the winds would have to be coherently directed so as to distort the isophotes in a manner similar to that which occurs naturally by merging. Recent simulations incorporating winds in $\Omega = 1$ clusters show little effect on the overall morphology (Metzler in this proceedings). Adding radiative cooling would produce a large central cooling flow, but there is no reason to suspect this will strongly affect the morphology of the outer regions. Finally, unrealistically large tidal torques would be required to distort the structure of the X–ray gas in the inner $\sim 1$ Mpc region, where the bulk of the X–rays are observed.

It all boils down to this. Generating anisotropy in the gas distribution at late times requires a directed source of energy input with magnitude comparable to the binding energy of the cluster. The most natural source for such directed energy is the merging of two systems of roughly comparable mass. To save the low $\Omega_o$ models, one needs to come up with a mechanism(s) which replaces merging, but produces the same effects. It is not at all clear how to do this.

## 4. X–ray Binding Mass Estimates

We have used the above simulations to test the accuracy of estimating the binding mass of clusters using the assumptions of hydrostatic equilibrium and a 'beta–model' for the gas distribution [23]. The mass estimate can be written as

$$M(<r) = -\frac{kT(r)r}{\mu m_p G}\left[\frac{d\ln\rho}{d\ln r} + \frac{d\ln T}{d\ln r}\right]. \qquad (2)$$

If the gas is close to isothermal at temperature $T$ and the density follows the usual form of $\rho(r) = \rho_o(1 + (r/r_c)^2)^{-3\beta/2}$, this becomes

$$M_{bind}(<r) = \frac{kTr}{\mu m_p G}\,3\beta\,\frac{(r/r_c)^2}{1+(r/r_c)^2}. \qquad (3)$$

ASCA observations reported at this meeting by Ikebe and Mushotzky indicate that the run of temperature with radius is close to isothermal. This need not imply exact isothermality throughout the cluster; only the azimuthal average need not vary strongly with radius. Briel showed data for A2256 from ROSAT which indicates a spatially varying temperature consistent with that expected from a recent merger. The simulated clusters generally show spatial variation, though the degree is dependent on the recent dynamical history (see Fig. 5 below). When radially averaged, the temparature profiles are usually weakly falling functions of radius out to the shock front separating the infall from hydrostatic regimes. Beyond this radius, which occurs at a density contrast $\mathcal{O}(10^2)$, the mean radial temperature drops rapidly.

To test the accuracy of the above estimators, the simulations used in the preceeding section were 'imaged' at $z = 0.04$ along each principal spatial axis with a 7200 sec exposure in the ROSAT passband. A Poisson photon noise field was added and then the mean value subtracted from the image. The radially averaged surface brightness maps were then fit to the standard form $\Sigma_x(\theta) = \Sigma_o(1 + (\theta/\theta_c)^2)^{-3\beta_{fit}+1/2}$. The central, emission weighted temperature $T_e$ was similarly determined. The values of $\beta_{fit}$, $r_c$ (from $\theta_c$) and $T_e$ were then used in equation (3) to estimate the binding mass as a function of radius. A total of 24 images for each cosmology were so generated.

We find that, at small radii where the density contrast is $\gtrsim 10^4$, the mass is typically underestimated, by factors up to 3. It is tempting to apply this result to the discrepancy raised by recent mass determinations from strong lensing (see Miralde-Escudé, Soucail and Babul in this proceedings). However, it is not clear whether this result is caused by physical or numerical effects, since it close to the resolution limit of these experiments. Also, the core of real clusters is a complicated place, with complex dynamics involving cooling flows which is not modeled in these experiments. It is, I believe, premature to claim that X–ray mass estimates are systematically low by large factors in the cores of rich clusters. It remains a real possiblity which should be pursued with future, high resolution simulations.

Farther out in the cluster, where the density contrast is between $10^3$ and $10^2$, the isothermal mass estimate is rather accurate. Fig. 4 shows histograms of the ratio of estimated to true mass measured at radii where the density contrast is 300. For all three cosmological models, the estimate is typically accurate to within 50% at this density contrast. Similar results are obtained for estimates at fixed metric radii of 1 Mpc in the $\Omega=1$ models and 0.5 Mpc in the low density runs. (The low $\Omega_o$ clusters are more compact than their critical density counterparts.)

The upshot is that one should feel confident in trusting, at the 50% level, binding mass determinations based on the isothermal, hydrostatic model around radii of $0.5-1$ Mpc. Models incorporating ejection from early–type galaxies display similar behavior [19, 7]. As a consequence, the high baryon mass fraction in clusters [25] should not be construed as arising from a large factor underestimate in their binding masses.

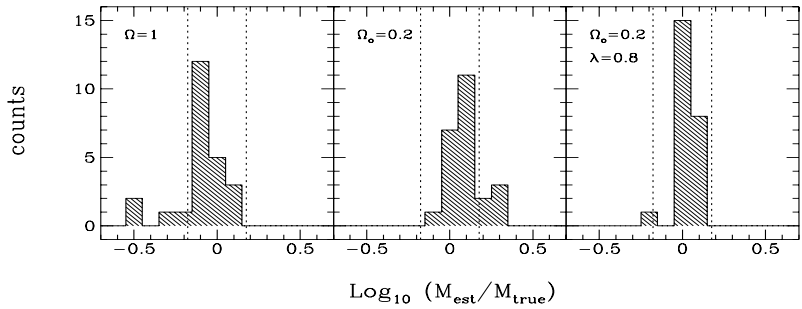

Figure 4. Histograms showing the decimal logarithm of the errors in isothermal binding mass estimates in a set of simulations. The estimates are at radii where the mean interior density contrast is 300, which is at a radius of of ∼ 1 Mpc for rich clusters. The estimates are typically accurate to within 50% (dashed vertical lines). The outliers in the $\Omega = 1$ runs are mergers in progress.

## 5. Dynamical Biasing of Galaxies in Clusters

It is well known that dynamical mass estimates based on the virial theorem applied to cluster galaxy kinematics have yielded mass to light ratios around a factor 5 smaller than that required to reach closure density [11]. The two possible interpretations are: (i) the estimate is unbiased and $\Omega_o \sim 0.2$ or (ii) there is a systematic bias in the estimate which makes it consistent with $\Omega = 1$. One possibility for the latter is that the dark matter is very weakly clustered on comoving scales $\lesssim 10$ Mpc. Another is that the galaxies are condensed toward the cluster center, and that one is measuring only some inner fraction of the total cluster mass and missing an extended, outer dark envelope. The latter issue has been investigated numerous times with N–body experiments over the past decade. Unfortunately, the interpretation of these simulations is clouded by the rather naive way in which galaxies were represented within the cluster.

Ideally, one would like to form galaxies *in situ* and subsequently follow their hierarchically clustering to the scale of rich clusters. Since local gas temperatures and densities are known, radiative cooling rates can be calculated (with or without photoionization heating). Gas which is sufficently dense can cool rapidly, lose pressure support and sink toward the bottom of the local potential well. The standard lore is that copius star formation ensues once the baryons become self–gravitating, creating in the process some sort of galactic unit.

Uncertainty in calculating cooling rates arises from lack of resolution — the baryonic matter associated with a single particle or cell, which is typically $\gtrsim 10^8 M_\odot$, is assumed to be a single phase medium characterised by a single density and temperature. At very high redshifts, this may be a reasonable approximation. However, in strongly non–linear clustered regions, a complex multi–phase medium akin to our own interstellar medium will develop. If a typical galaxy is resolved by enough particles or cells, a crude multi–phase will develop, with cold, dense knots one associates with star forming regions surrounded by halos of hot, rarefied gas.

Simulations of this sort have been successful recently in producing clusters with anywhere from three [16] to several tens [9] of such 'galaxies' within them. An example is shown in Fig. 5. This cluster was modeled with P3MSPH using $2 \times 64^3$ particles to represent the dark matter and baryons. The simulation modeled a periodic cube 22.5 Mpc on a side in a standard cold, dark matter universe. The limiting spatial resolution was ∼ 30 kpc and the mass per baryon particle was $3 \times 10^8 M_\odot$. An $L_*$ galaxy would thus be modeled by about 300 particles.

The cluster shown in Fig. 5 forms from a merger at $z \sim 0.3$ (second row) involving roughly three major components. Prior to this, the dark matter and baryons trace each other well on scales resolved in the figure. During the merger, jets of hot gas can be seen squirting out in directions perpendicular to the merger axis (compare the third with the first two columns of the second row). By $z = 0.1$, the hot gas has relaxed and is noticeably rounder than the dark matter distribution. A small group infalling from the lower right produces an X–ray emission feature and a fairly sharp temperature gradient in the cluster gas as the ram pressure of the infalling group compresses the gas intervening between it and the cluster core. The final X–ray structure of the cluster is quite regular, since no late mergers occur to stir up the gas. A mild, negative temperature gradient exists in the X–ray gas.

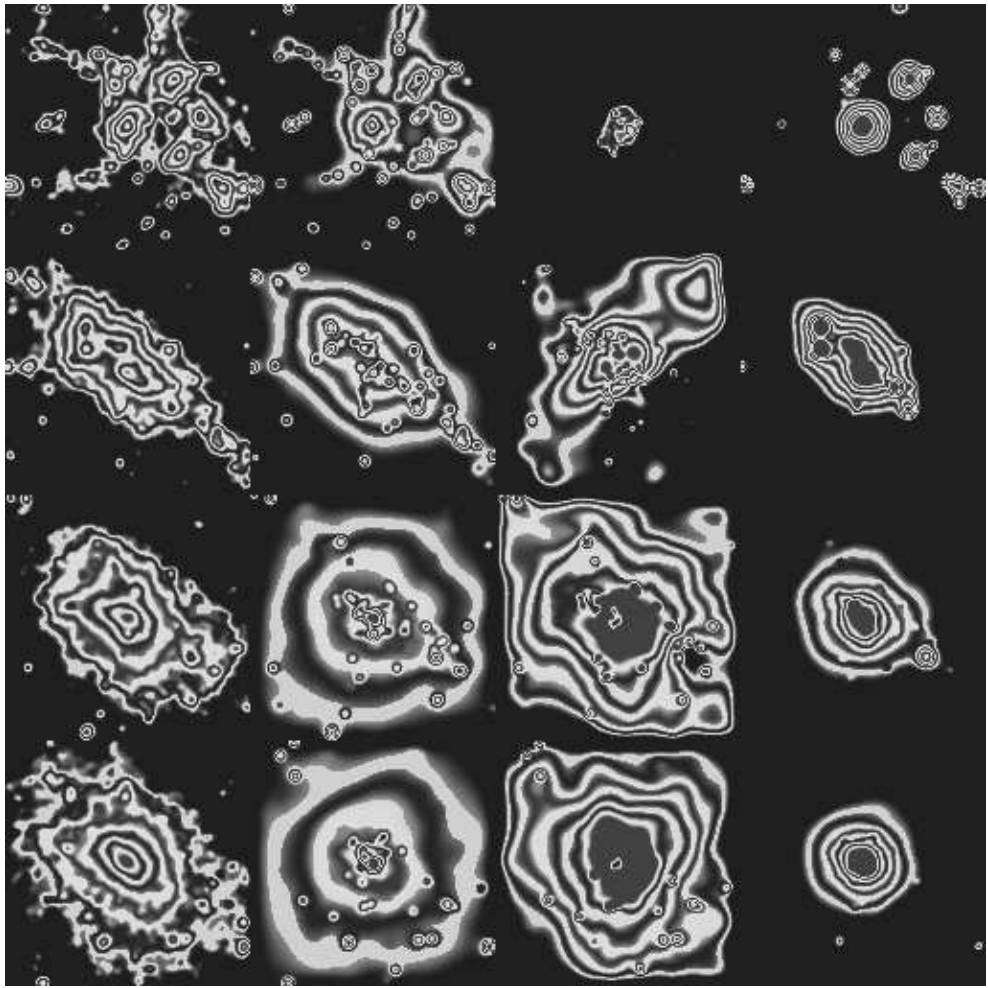

Figure 5. Formation of a cluster in a CDM universe (from Frenk, Evrard & White 1994). The columns show (left to right) projected dark matter density, projected gas density, emission weighted gas temperature and X–ray surface brightness in the ROSAT passband. The rows show the cluster at redshifts (top to bottom) $z = 0.7$, $0.3$, $0.1$ and $0$. Each panel shows a physical 3 Mpc region. The X–ray images were made by fixing the observer—cluster distance to be equivalent to $z = 0.03$. The spacing between light or dark bands is approximately a factor of two except for the temperature maps where the spacing is $\sim 25\%$. See the text for further discussion.

'Galaxies' are visible in the figure as small blips in the projected gas distribution. After the merger, the dark matter distribution has local density enhancements in the periphery of the cluster, but the central regions are relatively smooth. The central region in the baryons has much more structure due to the two–phase nature of the gas.

The galaxies represent a cooler and more concentrated population with respect to the dark matter. The ratio of velocity dispersions $\sigma_{gal}/\sigma_{DM}$ (the 'velocity bias' parameter) is $\sim 75 - 85\%$. The ratio of the half–mass radii $R_{gal}/R_{DM}$, determined from the known 3D positions and using 1.6 Mpc as an outer radius, shows a much more pronounced bias. Application of the virial theorem to determine binding masses results in a large (factor $2-4$) underestimate of the total mass of the cluster.

What is worrisome about this treatment is that the galaxies are assumed to be purely gaseous throughout the evolution of the cluster. Their interactions with the surrounding medium and with each other during collisions entail viscous drag, which is unphysical for a galaxy comprised mainly of stars. Of particular concern is the fact that the largest galaxy in the center of the cluster ends up containing *more than half* of the total baryons in cluster galaxies. Although bright, central cD galaxies are not uncommon in rich clusters, it is not the norm for the central cD to be brighter than the sum of all the other galaxies in the cluster.

To test the effects of this purely collisional treatment on the galactic dynamics within the cluster, we performed another run in which the SPH gas particles in galaxies at $z = 0.7$ were instantaneously turned into collisionless 'stars'. A collisionless, two–fluid run, using as initial conditions the dark matter particles and galaxies comprised of the star particles above, was then evolved from $z = 0.7$ to 0.

The galaxies in the collisionless treatment were generally more extended than their gas dynamic counterparts. Examination of galaxies' trajectories showed that the predominantly radial infalling orbits tended to take the galaxies very close to the cluster center on their first infall. In the gas dynamic case, the viscous gas interactions in the high density, central region braked the galaxies, effectively trapping them in the cluster core where they quickly merged with the central cD. In contrast, the galaxies comprised of collisionless stars flew through the center relatively undisturbed, as expected if the cluster velocity dispersion is larger than the internal velocities of the galaxies (which is the case here). No extremely large central galaxy formed. Instead, two galaxies of comparable mass separated by 0.5 Mpc were the most conspicuous objects in the cluster at $z = 0$. This situation is reminiscent in the Coma cluster.

There still remained some level of bias in the collisionless galaxies. A velocity bias of similar magnitude persisted, while the spatial distribution depended on the galaxy mass cutoff. The most massive galaxies were more concentrated than the dark matter while the set of all galaxies above a minimum 32 particle count ($10^{10} M_\odot$ in baryons) was spatially unbiased with respect to the dark matter. The different behavior of the two mass groups is not a transient result, since the same trend existed at earlier redshifts. It may be that the result is 'inherited' from the gas dynamic run via the initial conditions; the gravitational clustering being inefficient at erasing the memory of the initial bias. More details of these results are given elsewhere [5, 10].

To summarize, the issue of biases in the galaxy distribution within clusters remains uncertain. Observations of luminosity dependent clustering would be very helpful in constraining models.

## 6. Summary

Dynamical modeling of clusters of galaxies has improved significantly in the past few years, with the advent of simulation algorithms capable of handling the coupled evolution of multiple components representing dark matter, intracluster gas and galaxies. The new generation of experiments has opened up a new avenue for constraining $\Omega_o$ by using the X–ray morphology of rich clusters. They have also shed light on the accuracy of X–ray based binding mass estimates and the cluster baryon fraction inferred from them.

Issues which are more intimately linked to galaxy/star formation remain relatively poorly understood. Although definitive answers to the question of dynamical biases for galaxies in clusters remain elusive, results emerging from a variety of independent treatments suggest that galaxies should give a velocity dispersion estimate biased slightly ($10-30\%$) low with respect to the dark matter. Optical mass estimates are likely to underestimate the total binding mass, but the magnitude of this effect is fairly uncertain. A firmer understanding of the star formation history of galaxies is required to significantly advance beyond our present position.

This work was supported by a NATO International Travel Grant, NASA Theory Grant NAGW-2367 and NSF via supercomputer resources.